%% file: main.tex
\documentclass[conference]{IEEEtran}
\IEEEoverridecommandlockouts
\usepackage{cite}
\usepackage{amsmath,amssymb,amsfonts}
\usepackage{algorithmic}
\usepackage{graphicx}
\usepackage{textcomp}
\usepackage{xcolor}
\usepackage{listings}
\usepackage{caption}
\usepackage{booktabs} 
\usepackage{array}    
\usepackage{multirow}
\usepackage{tabularx}
\usepackage{pifont}
\usepackage{threeparttable}
\usepackage{fancyhdr} 
\usepackage{url}
\usepackage{hyperref}

\newcommand{\etc}{\textit{etc}.}

\def\BibTeX{{\rm B\kern-.05em{\sc i\kern-.025em b}\kern-.08em
    T\kern-.1667em\lower.7ex\hbox{E}\kern-.125emX}}

\begin{document}

\title{A Scenario-Oriented Benchmark for Assessing AIOps Algorithms in Microservice Management
}

\author{
    \IEEEauthorblockN{
    Yongqian Sun\IEEEauthorrefmark{1}, 
    Jiaju Wang\IEEEauthorrefmark{1}, 
    Zhengdan Li\IEEEauthorrefmark{1}, 
    Xiaohui Nie\IEEEauthorrefmark{2},
    Minghua Ma\IEEEauthorrefmark{5},
    Shenglin Zhang\IEEEauthorrefmark{1}, 
    Yuhe Ji\IEEEauthorrefmark{1}
    } 
    \IEEEauthorblockN{
    Lu Zhang\IEEEauthorrefmark{1},
    Wen Long\IEEEauthorrefmark{1},
    Hengmao Chen\IEEEauthorrefmark{4},
    Yongnan Luo\IEEEauthorrefmark{1},
    Dan Pei\IEEEauthorrefmark{3}
    }
    
    \IEEEauthorblockA{\IEEEauthorrefmark{1}\textit{Nankai University},
    \IEEEauthorrefmark{2}\textit{CNIC, CAS},
    \IEEEauthorrefmark{5}\textit{Microsoft}, \IEEEauthorrefmark{3}
    \textit{Tsinghua University},
    \IEEEauthorrefmark{4}\textit{BizSeer}
    }
    
    
} 

\newcommand{\workname}{\textit{MicroServo}}
\newcommand{\datasetname}{\textit{MicroServo}}

\maketitle
\thispagestyle{plain}
\pagestyle{plain}

\begin{abstract}
\input{sections/abstract}
\end{abstract}


\section{Introduction}
\label{sec:introduction}
\input{sections/introduction}

\section{Typical AIOps Tasks for Microservice Systems}
\label{sec:background}
\input{sections/background}

\section{System design}
\label{sec:system design}
\input{sections/system_design2}

\section{Case Study}
\label{sec:experiment}
\input{sections/experiment2}

\section{Conclusion and Future Prospects}
\label{sec:conclusion}
\input{sections/conclusion}

\bibliographystyle{unsrt}
\bibliography{bibfile}

\end{document}

%% file: sections/abstract.tex
AIOps algorithms play a crucial role in the maintenance of microservice systems. Many previous benchmarks' performance leaderboard provides valuable guidance for selecting appropriate algorithms. However, existing AIOps benchmarks mainly utilize offline datasets to evaluate algorithms. They cannot consistently evaluate the performance of algorithms using real-time datasets, and the operation scenarios for evaluation are static, which is insufficient for effective algorithm selection. To address these issues, we propose an evaluation-consistent and scenario-oriented evaluation framework named \workname. 
The core idea is to build a live microservice benchmark to generate real-time datasets and consistently simulate the specific operation scenarios on it. \workname\ supports different leaderboards by selecting specific algorithms and datasets according to the operation scenarios. It also supports the deployment of various types of algorithms, enabling algorithms hot-plugging. At last, we test \workname\ with three typical microservice operation scenarios to demonstrate its efficiency and usability.

%% file: sections/introduction.tex
Microservice architecture, which utilizes loosely coupled and independently scalable services, has revolutionized the development and deployment of web applications\cite{microservice-architecture}. This architectural model is widely adopted due to its ability to enhance agility, improve scalability, and facilitate continuous deployment, making it ideal for businesses that need to update and scale components frequently. However, due to the complex structure of microservice systems, relying solely on manual efforts to address failures is insufficient. Therefore, researchers have utilized three types of data that reflect system states: logs, metrics, and traces, to develop many AIOps algorithms \cite{diagfusion,cloudrca,timesnet,usad,logcluster,traceanomaly,microcbr} for the maintenance of the systems. 

Microservice management consists of several tasks, including anomaly detection, root cause localization, failure classification, and \etc~The selection and evaluation of algorithms for these tasks are crucial. This paper defines an \textit{operation scenario} as the fault orchestration within the dataset. For example, if an evaluator wants to assess the performance of anomaly detection algorithms under network faults, the operation scenario for this evaluation is network faults. Alternatively, if an evaluator wants to evaluate the performance of anomaly detection algorithms in a real production environment, which may encounter various faults, the operation scenario is a combination of different faults. 

The Current standard process for evaluating an algorithm typically involves identifying a dataset that suits the specific operation scenario, selecting multiple baseline algorithms, reproducing these algorithms, and finally comparing their performance on the dataset using representative evaluation metrics. Among these steps, data collection, data cleaning, and reproducing baseline algorithms are notably tedious and time-consuming. Furthermore, selecting the appropriate algorithm poses certain complexities, as different groups have varying algorithmic needs. For individuals not well-versed in algorithms, determining whether to select supervised or unsupervised methods, deep learning, reinforcement learning, or other approaches is particularly challenging.

In light of these issues, previous researchers have developed benchmarks for evaluating algorithms, such as timeseriesbench\cite{timeseriesbench}. These benchmarks use offline datasets, reproduce several algorithms, and compile the experimental results into evaluation leaderboards. However, they cannot consistently evaluate the performance of algorithms using read-time datasets. And since they use offline datasets, their evaluation leaderboards are only valuable for algorithm evaluation in fixed operation scenarios. Therefore, we aim to implement an evaluation-consistent and scenario-oriented evaluation framework. This framework first simulates operation scenarios on a live microservice benchmark, then uses real-time datasets for algorithm evaluation, while also maintaining a repository of open-source algorithms for evaluators to use.

Implementing the above evaluation framework faces the following challenges: 1) Maintaining different algorithms on the platform may encounter dependency environments conflicts. 2) Consistently evaluate algorithms calls for a live microservice benchmark. Scenario-oriented evaluation means that each evaluation requires using datasets corresponding to the specific operation scenarios. The entire process, from fault injection, data collection and export, to using the data for evaluation, is labor-intensive. Therefore, automation is necessary.

Based on these challenges, we provide the industry with a scenario-oriented, evaluation-consistent evaluation platform \workname, offering targeted and practical evaluation solutions. \workname\ has two main features: 1) \textbf{Algorithm Containerization}: We develop an algorithm hot-plugging function. By deploying algorithms in containers, we achieve one-click deployment and removal, perfectly resolving the environment conflicts. 2) \textbf{Process Automation}: We build a complete ecosystem of system-tools-applications based on a live microservice benchmark, ensuring the real-time support. With fault orchestration, the platform automatically injects faults into the microservice system, collects datasets, and provides them for evaluation applications, which eliminates the limitations of operation scenarios of offline datasets and saves manual time costs.

The paper's main contributions are as follows:
\begin{enumerate}
    \item We develop a comprehensive framework \workname~\footnote{The source code is available at \href{https://github.com/MicroServo/microservo}{https://github.com/MicroServo/microservo}, and the system is deployed and can be accessed by \href{https://microservo.aiops.cn}{https://microservo.aiops.cn}.}. \workname\ facilitates fault injection, data collection and the upload and execution of various types of algorithms. It also deploys an evaluation application that can use real-time datasets and select algorithms integrated into the platform for evaluation, then display the evaluation leaderboard.
    \item We propose a simple dataset format specification for microservice systems, allowing algorithms using this format to integrate into the platform.
    \item We design a scenario-oriented, evaluation-consistent evaluation solution, which enables consistently evaluating specific operation scenarios thus more practical.
    \item We release datasets collected from \workname \footnote{Dataset available at: https://github.com/MicroServo/dataset} that include typical operation scenarios. Our commitment to maintaining and updating these datasets supports ongoing research and development efforts in the field.
\end{enumerate}

%% file: sections/background.tex
\subsection{Anomaly Detection}
Anomalies, such as unexpected fluctuations or rapid deviations from normal patterns, often indicate potential faults, including hardware crashes, service disruptions, and software bugs. The primary objective of anomaly detection is to identify anomalous behavior in both system status and user behavior to prevent system crashes and mitigate potential disruptions to the business. Anomaly detection typically consists of two stages: offline model training and online anomaly detection. In the offline model training stage, sufficient data is used to train an anomaly detection model. The trained model outputs an anomaly score for each time point during the online detection stage, indicating the likelihood of being anomalous\cite{anomaly-empirical}.

\subsection{Failure Diagnosis}
Further diagnostic procedures are undertaken to ascertain the potential root causes upon detecting an anomaly within a microservice system. This diagnostic process primarily focuses on two key aspects: root cause localization and failure classification.

\textit{Root Cause Localization}. Existing methodologies predominantly utilize Key Performance Indicators (KPIs), application logs, and distributed traces to analyze the root causes. Some approaches directly process these data sources to identify which services are experiencing anomalies. Others utilize this data to construct graphs or topological models that depict the dependencies among services within the system\cite{failure-diagnosis}.

\textit{Failure Classification}. This process involves studying the patterns or fingerprints\cite{microcbr} of past failure cases. When a new anomaly is encountered, the system compares it with the learned patterns, thereby categorizing it into a known failure type or identifying it as a new type.

%% file: sections/system_design2.tex
\begin{figure}[ht]
    \centering
    \includegraphics[width=1.0\linewidth, height=0.20\textheight]{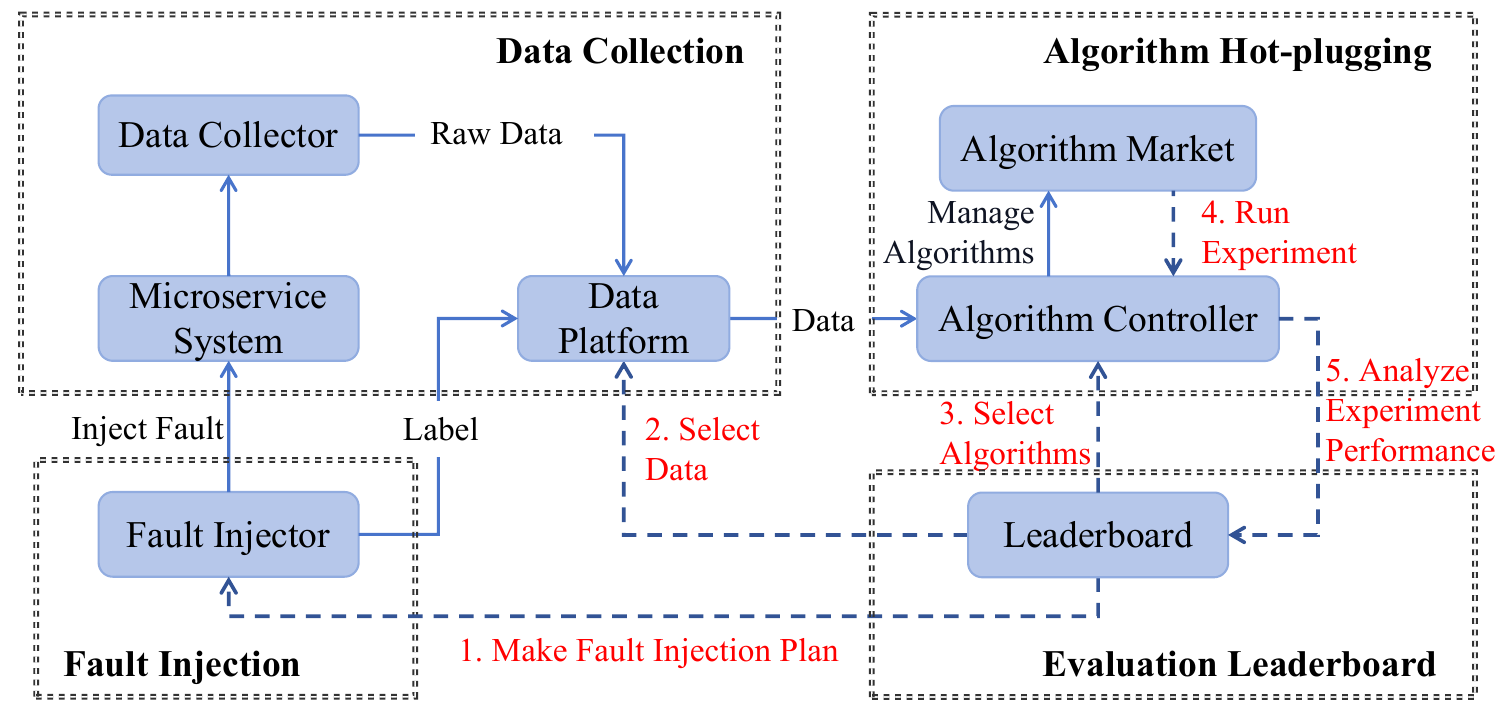}
    \caption{Framework of \workname.}
    \label{fig:platform-framework}
\end{figure}

\workname\ is a user-friendly and highly extensible platform primarily designed to evaluate the performance of AIOps algorithms. Our focus is on two main aspects: how to support the execution of different types of algorithms and how to conduct scenario-oriented evaluations. Fig.\ref{fig:platform-framework} illustrates the architecture of \workname, which comprises four main modules. Each module is responsible for a critical aspect of the evaluation process, functioning cooperatively yet independently.

\begin{enumerate}
    \item \textit{Fault Injection}: The fault injection module is capable of simulating operation scenarios in real production environments and offers an intuitive way for fault orchestration.
    \item \textit{Data Collection}: To be compatible with different types of algorithms, we propose a simple dataset format under a microservice system, standardizing the data input format for the platform's algorithms.
    \item \textit{Algorithm Hot-plugging}: This module is responsible for managing all the algorithms integrated into the platform. An algorithm can be quickly integrated into the platform with a simple operation. Algorithms on the platform also support one-click removal.
    \item \textit{Evaluation Leaderboard}: We design a consistent scenario-oriented evaluation mechanism. By specifying datasets and selecting relevant algorithms, a new evaluation leaderboard can be easily created.
\end{enumerate}

\subsection{Fault Injection}
\label{sec:fault-injection}
Our injection design selects a calendar format to provide an intuitive view of fault orchestrations, detailing when and where each fault will occur. We also streamline the fault injection process to simplify user interactions, making planning and executing chaos experiments more straightforward and less time-consuming. As listed in Table \ref{table:fault_injection}, \workname\ currently supports five main fault types. \workname\ leverages ChaosMesh\cite{chaosmesh} to inject the above faults into the microservice system. ChaosMesh is a chaos engineering platform for Kubernetes\cite{kubernetes2024}, designed to experiment on a distributed system to build confidence in the system’s capability to withstand turbulent and unexpected conditions in production. In the future, we will continue to add other types of faults supported by Chaos Mesh into the fault injector, such as IO-Type and HTTP-Type.

\input{tables/chaos_mesh}

\workname\ supports both immediate and scheduled fault injection methods. The fault injection process involves the injector creating a chaos description object of ChaosMesh and committing it to Kubernetes. Kubernetes then identifies and passes these fault configurations to the ChaosMesh controller containers. Finally, the controller containers handle the actual fault injection logic. Our fault injector takes fault definition as its input and injects the defined faults at proper timing. A fault definition contains the fault description. Each fault description includes information on the fault target instance, fault start time (when to inject this fault), fault duration (how long this fault lasts after being created), and fault behaviors dictionary. Each key in fault behaviors corresponds to a fault type, and its value is the detailed configuration (i.e., parameters) of that type of fault.

\subsection{Data Collection}
\label{sec:data-collection}
To generate a realistic online microservice system dataset, we deploy the open-source Online Boutique\cite{onlineboutique} provided by GoogleCloudPlatform. Using three collectors, we gather metrics, logs, and trace data, storing them in the data platform. The data is then cleansed according to a specified format, and data export APIs are provided for algorithms integrated into the platform.

In practice, \workname\ uses Prometheus\cite{prometheus2024} as the metric collector, Filebeat\cite{elasticbeats2024} as the log collector, and Elastic APM\cite{elasticapm2024} as the trace collector. Prometheus is a popular monitoring system for distributed data endpoints and a time-series database capable of storing collected metrics. Filebeat is a lightweight log data collector that reads data from specified log files and sends it to backend systems such as Elasticsearch\cite{elasticsearch2024} or Logstash\cite{logstash2024} for further processing and storage. Elastic APM is a tool for monitoring the performance of distributed applications, capable of collecting and analyzing application trace data in real-time.

To enable different algorithms to use the platform's dataset for experiments, we establish a simple dataset format specification, defining the format for each type of data. The metric data includes 17 container-level metrics and 10 service-level metrics, each stored in separate CSV files. The format for these files is as follows: (\textit{timestamp}, \textit{cmdb\_id}, \textit{kpi\_name}, \textit{value}), where each line represents a value sampled from one instance of a service. The log data are preserved in a CSV file. Each row in the CSV file corresponds to an individual log entry, representing a singular event or state captured by the microservice's logging system. The format of the CSV file is composed of the following columns: (\textit{log\_id}, \textit{timestamp}, \textit{date}, \textit{cmdb\_id}, \textit{message}). The traces are also stored in a CSV format. The CSV file is structured with columns that consist of: (\textit{timestamp}, \textit{cmdb\_id}, \textit{parent\_span}, \textit{span\_id}, \textit{trace\_id}, \textit{duration}, \textit{type}, \textit{status\_code}, \textit{operation\_name}). By standardizing the dataset format, we have unified the input format for the algorithms to be introduced, thereby laying the foundation for the extensibility of algorithms on the platform.

\subsection{Algorithm Hot-plugging}
\label{sec:algorithm-hot-plugging}
\workname\ implements one-click startup and removal of algorithms. Excluding the time required for image building, starting or deleting a container only takes a minimal amount of time, achieving true hot-swapping of algorithms. Moreover, \workname\ is compatible with different types of algorithms. Currently, we have deployed six anomaly detection algorithms, three root cause localization algorithms, and four fault classification algorithms, all of which are open-source and frequently used as baselines for comparison. We commit to continuously adding new algorithm types and open-source algorithms into \workname.

The algorithm hot-plugging module consists of two main parts: Algorithm Market and Algorithm Controller. Algorithm Market, as the name suggests, algorithms are deployed on the platform, like goods in a market for evaluators to choose from. Considering the different environment dependencies of various algorithms, we deploy each algorithm in a separate Docker container\cite{dockerengine2024}. This approach allows for easy addition or removal of algorithms by simply operating on the corresponding container, which aligns well with the design principle of decoupling. The Algorithm Controller manages the starting, restarting, and deleting of containers and oversees the experiments conducted by the algorithms.

\begin{figure}[htbp]
    \centering
    \includegraphics[width=0.30\linewidth, height=0.15\textheight]{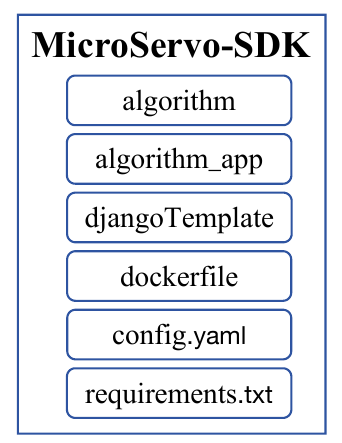}
    \caption{Structure of MicroServo-SDK.}
    \label{fig:microservo-sdk}
\end{figure}

\begin{figure*}[htbp]
    \centering
    \includegraphics[width=1.0\linewidth, height=0.25\textheight]{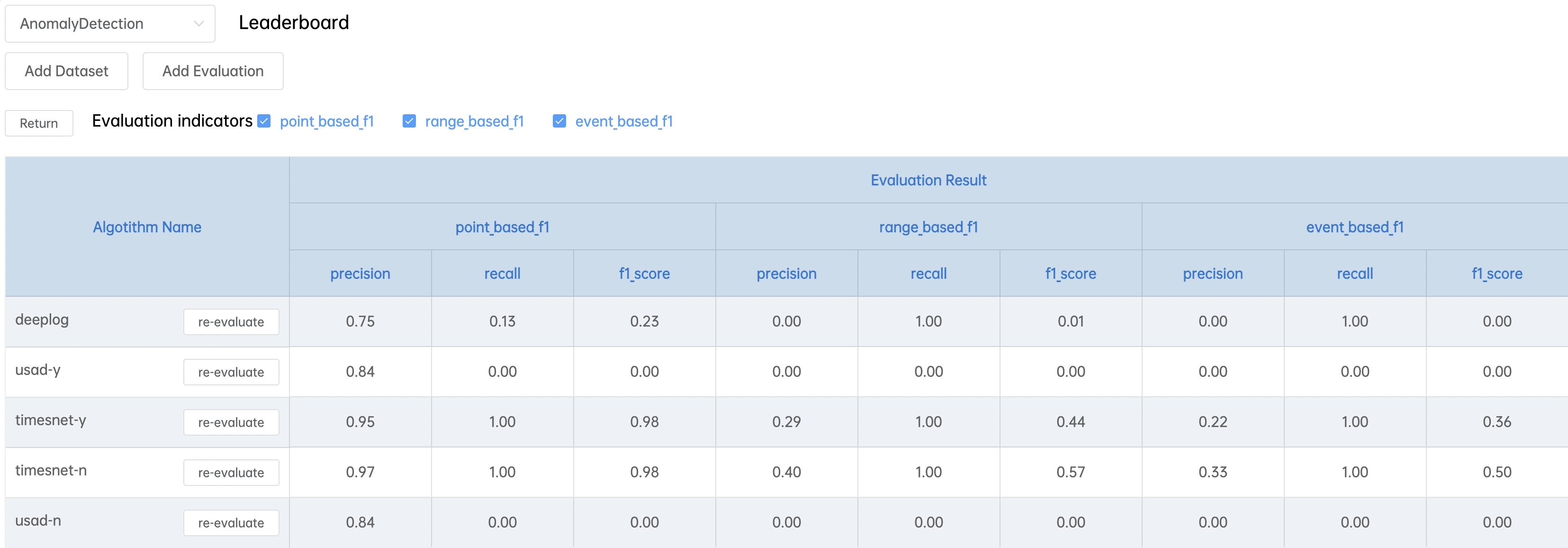}
    \caption{Example of Evaluation Leaderboard.}
    \label{fig:leaderboard}
\end{figure*}
In order to ease the procedure of uploading algorithms, we design Microservo-SDK\footnote{We design a template called MicroServo-SDK for algorithm hot-plugging, available at https://github.com/MicroServo/hot-plugging}, a template for Uploading. The structure of MicroServo-SDK is illustrated in Fig.\ref{fig:microservo-sdk}. Users need to place the algorithm code under the algorithm directory and implement the training and detection methods of the Algorithm class in algorithm\_app/algorithm.py. Finally, by adding the algorithm's dependencies to requirements.txt, the algorithm can be uploaded to the platform for deployment.

\subsection{Evaluation Leaderboard}
\label{sec:leader-board}
To assist evaluators in assessing the best-performing algorithms in specific operation scenario, we design a consistent and dynamic scenario-oriented evaluation mechanism and proposed a five-stage evaluation process, as illustrated in Fig.\ref{fig:platform-framework}: 1) Make a fault injection plan and inject it into the microservice system. 2) Select the dataset for the corresponding time period. 3) Choose the algorithms to be evaluated on the platform. 4) Conduct experiments. 5) Analyze the experiment results.

We provide scenario-oriented evaluations because algorithms with low scores in the leaderboard may perform better in other scenarios. Our evaluation approach supports both complex fault orchestration and a specific fault type, allowing evaluators to identify algorithms best suited to their particular needs. Consistency is reflected in the usage of real-time datasets. As for dynamic, selecting several algorithms for evaluation does not mark the end of the evaluation process. By retaining the datasets from each evaluation, evaluators can add other algorithms, including their own, at any time. The results of the newly added algorithms also update the current evaluation leaderboard. This module's primary function is to support the platform in executing various types of algorithms and to manage their deployment and experimentation.

Fig.\ref{fig:leaderboard} shows an evaluation leaderboard for anomaly detection in \workname. We construct an operation scenario for anomaly detection of network faults. Therefore, we select a dataset that only contains network faults and choose several anomaly detection algorithms—TimesNet\cite{timesnet}, DeepLog\cite{deeplog} and USAD\cite{usad}—for evaluation. By calculating evaluation metrics from the assessment results, we generate an evaluation leaderboard. The leaderboard shows that TimesNet achieves an F1-score significantly higher than the other algorithms. We provide three common evaluation metrics for anomaly detection algorithms: Point-based\cite{timeseriesbench}, Range-based\cite{timeseriesbench}, Event-based\cite{timeseriesbench}, three evaluation metrics for fault classification: \textit{Accuracy@k}\cite{39,64,65,cloudrca,76}, \textit{Avg@k}\cite{39,64,65,cloudrca,76}(Average Accuracy), \textit{MAR}\cite{70,118,119,143}(Mean Average Rank) and three evaluation metrics for root cause localization: \textit{Micro-F1}\cite{44,47,81,150,cloudrca}, \textit{Macro-F1}\cite{cloudrca}, \textit{Weighted-F1}\cite{81}. We will continue to update these evaluation metrics, striving to provide a more comprehensive set of evaluation standards.

%% file: tables/chaos_mesh.tex
\begin{table}
    \centering
    \caption{Supported Types of Fault Injections}
    \resizebox{\linewidth}{!}{
        \begin{tabular}{llll|llllll}
        \toprule
        \textbf{Name} & \textbf{Description} \\
        \midrule
        CPU Stress & Occupies CPU to simulate high load conditions. \\
        Memory Stress & Consumes memory to simulate high usage conditions. \\
        Pod Failure & Disrupts a specified Pod to create a period of unavailability. \\
        Network Delay & Induces latency to mimic slow network conditions between containers. \\
        Network Loss & Emulates network packet loss between containers. \\
        \bottomrule
        \end{tabular}
    }
    \label{table:fault_injection}
\end{table}

%% file: sections/experiment2.tex
In this section, We first evaluate the performance of three types algorithms, then assess the effectiveness of some anomaly detection algorithms using two single-type scenarios. The experiments are conducted on a server with the following specifications: Intel(R) Xeon(R) CPU E5-2650, 2.20GHz, 125GB RAM, 7TB hard drive, running Ubuntu 20.04.2 LTS.

\input{tables/fault_affection}

\input{tables/root-cause-localization}
\subsection{Dataset}
We select three datasets to detect algorithms: \textit{Scenario 1}, \textit{Scenario 2} and \textit{Scenario 3}. \textit{Scenario 1} simulates various fault scenarios. \textit{Scenario 2} simulates CPU and memory stress faults. \textit{Scenario 3} simulates pod faults. \textit{Scenario 1} collects five days of data, containing 226 fault cases. \textit{Scenario 2}, \textit{Scenario 3} collects one days of data, containing 12 fault cases. The data types include metrics, logs, and trace data. The metrics collection interval is 15 seconds. We also analyze  the fault performance of each dataset, as shown in Table \ref{table:fault_affection}. Fault data in \textit{Scenario 2} may not be clearly reflected in the logs and traces, whereas \textit{Scenario 3} is more likely to be prominently reflected in the logs.

\input{tables/anomaly-detection}

\input{tables/failure-classification}

\subsection{Algorithms}
For evaluation on \textit{Scenario 1}, in failure classification algorithms, we select Diagfusion\cite{diagfusion}-FC, LogCluster\cite{logcluster}, MicroCBR\cite{microcbr} and Cloudrca\cite{cloudrca}. In root cause localization algorithms, we select Diagfusion-RCL, Automap\cite{automap}, TraceRCA\cite{tracerca}. And we select  JumpStarter\cite{jumpstarter}, NeuralLog\cite{neurallog}, TraceAnomaly\cite{traceanomaly} and TimesNet\cite{timesnet} for anomaly detection. For evaluation on \textit{Scenario 2} and \textit{Scenario 3}, we select USAD\cite{usad}, DeepLog\cite{deeplog} and TimesNet\cite{timesnet}.

\input{tables/ad-case-2}

\input{tables/ad-case-3}
\subsection{Results and Analysis}
Tables \ref{tab:root-cause-localization} \ref{tab:anomaly-detection} \ref{tab:failure-classification} present the leaderboards of three types of algorithms on \textit{Scenario 1}. In Table \ref{tab:anomaly-detection}, TimesNet performs relatively well but is not outstanding. The primary reason is that \textit{Scenario 1} includes various faults such as network faults, pod faults, and stress faults, each exhibiting different characteristics across the three types of data. For instance, network and stress faults are not clearly reflected in the logs, while pod faults are not apparent in the metrics and trace data. The algorithms evaluated are all unimodal, limiting their ability to detect anomalies comprehensively. In Table \ref{tab:failure-classification}, DiagFusion slightly outperforms CloudRCA. Both DiagFusion and CloudRCA are multimodal algorithms, capable of fully leveraging information from different data modalities. In Table \ref{tab:root-cause-localization}, DiagFusion also achieves the best results. In contrast, AutoMap performs very poorly, likely due to its reliance solely on metrics to construct a PC network for random walks.

The experimental results shown in Table \ref{tab:ad-case-2} and Table \ref{tab:ad-case-3} demonstrate that our scenario-oriented evaluation method is meaningful. Table \ref{tab:ad-case-2} presents the performance of anomaly detection algorithms under stress fault scenario, where DeepLog performs very poorly because stress faults are not reflected in log data. In Table \ref{tab:ad-case-3}, the scenario is changed to pod faults, which manifest as entries such as 'pod unable to connect' in the logs. Consequently, DeepLog's performance improves. On the other hand, TimesNet performs excellently in both scenarios because both faults are reflected in the metrics data.

In production environments, it is not always necessary for an algorithm to perform well in all scenarios. Often, evaluators may only need to find the best-performing algorithm for a specific scenario. Our proposed scenario-oriented evaluation method effectively assists evaluators in making such selections.

%% file: tables/fault_affection.tex
\begin{table}
    \centering
    \caption{Manifestation of Different Cases Across Data Categories}
    \resizebox{\linewidth}{!}{
        \begin{tabular}{lccc}
        \toprule
        \textbf{Name} & \textbf{Shown in Metrics} & \textbf{Shown in Logs} & \textbf{Shown in Traces} \\
        \midrule
        \textit{Case 1} & \ding{51} & \ding{51} & \ding{51} \\
        \textit{Case 2} & \ding{51} & \ding{55} & \ding{55} \\
        \textit{Case 3} & \ding{51} & \ding{51} & \ding{55} \\
        \bottomrule
        \end{tabular}
    }
    \label{table:fault_affection}
\end{table}

%% file: tables/root-cause-localization.tex
\begin{table*}[t]
\centering
\caption{Root Cause Localization Leaderboard on \textit{Scenario 1}}
\label{tab:root-cause-localization}
\resizebox{\linewidth}{!}{
    \begin{tabular}{lcccccccccc}
    \toprule
    & \multicolumn{10}{c}{\textbf{Evaluation Metrics}} \\ \cmidrule(l){2-11}
    \textbf{Algorithm} & \multicolumn{5}{c}{\textbf{Accuracy@k}} & \multicolumn{5}{c}{\textbf{Avg@k}} \\ \cmidrule(l){2-11}
    & \textbf{Accuracy@1} & \textbf{Accuracy@2} & \textbf{Accuracy@3} & \textbf{Accuracy@4} & \textbf{Accuracy@5} & \textbf{Avg@1} & \textbf{Avg@2} & \textbf{Avg@3} & \textbf{Avg@4} & \textbf{Avg@5} \\ \midrule
    Diagfusion\cite{diagfusion}RCL & 0.52 & 0.84 & 0.91 & 0.92 & 0.93 & 0.52 & 0.68 & 0.76 & 0.80 & 0.82 \\
    Automap\cite{automap} & 0.01 & 0.06 & 0.07 & 0.10 & 0.16 & 0.01 & 0.04 & 0.05 & 0.06 & 0.08 \\
    TraceRCA\cite{tracerca} & 0.27 & 0.31 & 0.48 & 0.55 & 0.68 & 0.27 & 0.29 & 0.35 & 0.40 & 0.46 \\ \bottomrule
    \end{tabular}
}
\end{table*}

%% file: tables/anomaly-detection.tex
\begin{table}[b]
\centering
\caption{Anomaly Detection Leaderboard on \textit{Scenario 1}}
\label{tab:anomaly-detection}
\resizebox{\linewidth}{!}{
    \begin{tabular}{lccccccccc}
    \toprule
    & \multicolumn{9}{c}{\textbf{Evaluation Metrics}} \\ \cmidrule(l){2-10} 
    \textbf{Algorithm} & \multicolumn{3}{c}{\textbf{Point-based}} & \multicolumn{3}{c}{\textbf{Range-based}} & \multicolumn{3}{c}{\textbf{Event-based}} \\ \cmidrule(l){2-10} 
     & \textbf{Precision} & \textbf{Recall} & \textbf{F1-score} & \textbf{Precision} & \textbf{Recall} & \textbf{F1-score} & \textbf{Precision} & \textbf{Recall} & \textbf{F1-score} \\ \midrule
    JumpStarter\cite{jumpstarter}    &   0.51    &   0.81   &    0.62   &   -   &   -   &  -    &   -   &  -  &   -   \\
    NeuralLog\cite{neurallog}    &   0.43    &   1   &   0.6   &  -  &   -   &  -    &   -   &  -  &   -   \\
    TraceAnomaly\cite{traceanomaly} &   0.37   &   0.56    &    0.45     &   -     &    -   &    -     &   -    &    -   &    -   \\
    TimesNet\cite{timesnet} &    0.95   &  0.65   &     0.77    &     0.44   &  0.67    &    0.53     &   0.44     &   0.67    &     0.53    \\\bottomrule
    \end{tabular}
}
\end{table}

%% file: tables/failure-classification.tex
\begin{table}[b]
\centering
\caption{Failure Classification Leaderboard on \textit{Scenario 1}}
\label{tab:failure-classification}
\resizebox{\linewidth}{!}{
    \begin{tabular}{lccccccccc}
    \toprule
    & \multicolumn{9}{c}{\textbf{Evaluation Metrics}} \\ \cmidrule(l){2-10} 
    \textbf{Algorithm} & \multicolumn{3}{c}{\textbf{Micro-F1}} & \multicolumn{3}{c}{\textbf{Macro-F1}} & \multicolumn{3}{c}{\textbf{Weighted-F1}} \\ \cmidrule(l){2-10} 
     & \textbf{Precision} & \textbf{Recall} & \textbf{F1-score} & \textbf{Precision} & \textbf{Recall} & \textbf{F1-score} & \textbf{Precision} & \textbf{Recall} & \textbf{F1-score} \\ \midrule
    LogCluster\cite{logcluster} &    0.60    &   0.60    &     0.60    &   0.49     &   0.48    &    0.43     &    0.75    &  0.60    &   0.63   \\
    Diagfusion\cite{diagfusion}FC &   0.70   &    0.70   &     0.70    &     0.51   &   0.45   &     0.47    &    0.77    &   0.70    &    0.73     \\
    MicroCBR\cite{microcbr} &   0.00     &    0.00   &     0.00    &     0.39   &   0.47   &     0.40    &    0.63    &   0.80    &    0.65     \\
    CloudRCA\cite{cloudrca}  &     0.67   &   0.67    &   0.67      &   0.33     &  0.50     &   0.40      &    0.44    &   0.67    &   0.53      \\ \bottomrule
    \end{tabular}
}
\end{table}

%% file: tables/ad-case-2.tex
\begin{table}
\centering
\caption{Anomaly Detection Leaderboard on \textit{Scenario 2}}
\label{tab:ad-case-2}
\resizebox{\linewidth}{!}{
    \begin{tabular}{lccccccccc}
    \toprule
    & \multicolumn{9}{c}{\textbf{Evaluation Metrics}} \\ \cmidrule(l){2-10} 
    \textbf{Algorithm} & \multicolumn{3}{c}{\textbf{Point-based}} & \multicolumn{3}{c}{\textbf{Range-based}} & \multicolumn{3}{c}{\textbf{Event-based}} \\ \cmidrule(l){2-10} 
     & \textbf{Precision} & \textbf{Recall} & \textbf{F1-score} & \textbf{Precision} & \textbf{Recall} & \textbf{F1-score} & \textbf{Precision} & \textbf{Recall} & \textbf{F1-score} \\ \midrule
    DeepLog\cite{deeplog}    &   0.75    &   0.13   &    0.23   &   -   &   -   &  -  & -   &  -  &  -  \\
    USAD\cite{usad}    &   0.88    &   0.33   &   0.48   &  0.50  &   0.33   &  0.40    &   0.50   &  0.33  &   0.40   \\
    TimesNet\cite{timesnet} &    0.90   & 1.00   &  0.95 &    0.30   &  1.00    &    0.46     &   0.25     &   1.00   &     0.40    \\\bottomrule
    \end{tabular}
}
\end{table}

%% file: tables/ad-case-3.tex
\begin{table}
\centering
\caption{Anomaly Detection Leaderboard on \textit{Scenario 3}}
\label{tab:ad-case-3}
\resizebox{\linewidth}{!}{
    \begin{tabular}{lccccccccc}
    \toprule
    & \multicolumn{9}{c}{\textbf{Evaluation Metrics}} \\ \cmidrule(l){2-10} 
    \textbf{Algorithm} & \multicolumn{3}{c}{\textbf{Point-based}} & \multicolumn{3}{c}{\textbf{Range-based}} & \multicolumn{3}{c}{\textbf{Event-based}} \\ \cmidrule(l){2-10} 
     & \textbf{Precision} & \textbf{Recall} & \textbf{F1-score} & \textbf{Precision} & \textbf{Recall} & \textbf{F1-score} & \textbf{Precision} & \textbf{Recall} & \textbf{F1-score} \\ \midrule
    DeepLog\cite{deeplog}    &   0.70    &   0.33   &    0.45   &   -   &   -   &  -   & -  &  -  &   - \\
    USAD\cite{usad}    &   0.84    &   0.00   &   0.00   &  0.00  &   0.00   &  0.00    &   0.00   &  0.00  &   0.00   \\
    TimesNet\cite{timesnet} &    0.95   & 1.00   &  0.98 &    0.29   &  1.00    &    0.44     &   0.22     &   1.00   &     0.36    \\\bottomrule
    \end{tabular}
}
\end{table}

%% file: sections/conclusion.tex
In this paper, we propose \workname, an algorithm-dynamic and scenario-oriented algorithm evaluation framework. To provide evaluations tailored to specific operation scenarios, \workname\ automates processes from fault injection, data collection and cleansing, to algorithm evaluation. Additionally, \workname\ supports deploying various algorithms, enabling one-click deployment and removal. To demonstrate the functionality of scenario-oriented evaluations, we conducted evaluations in three operation scenarios, presenting the corresponding evaluation leaderboards. In future work, we commit to deploying more state-of-the-art open-source algorithms on the platform. Additionally, we plan to add the capability to upload offline datasets, further enriching the evaluation functionality.